\documentclass{aa}
\usepackage{txfonts}
\usepackage{graphicx}
\usepackage{amssymb}
\usepackage{natbib}
\bibpunct{(}{)}{;}{a}{}{,}
\usepackage{color}

\def\BI  {(B-I)}
\def\VI  {(V-I)}
\def\VK  {(V-K)}

\def\Rc  {R_\mathrm{c}}
\def\Ic  {I_\mathrm{c}}
\def\Ks  {K_\mathrm{s}}

\def\Wbi {W_\mathrm{bi}}
\def\Wvi {W_\mathrm{vi}}
\def\Rb  {R_\mathrm{B}}
\def\Rv  {R_\mathrm{V}}

\def\EBV {E(B-V)}

\def\Mb  {M_\mathrm{B}}
\def\Mv  {M_\mathrm{V}}
\def\Mrc {M_\mathrm{Rc}}
\def\Mic {M_\mathrm{Ic}}
\def\Mwbi {M_\mathrm{Wbi}}
\def\Mwvi {M_\mathrm{Wvi}}
\def\Mj  {M_\mathrm{J}}
\def\Mh  {M_\mathrm{H}}
\def\Mks {M_\mathrm{Ks}}

\newcommand\Sec[1]  {Sec.~\ref{#1}}

\newcommand\Fig[1]  {Fig.~\ref{#1}}
\newcommand\Tab[1]  {Tab.~\ref{#1}}
\begin{document}
  \title{A new calibration of Galactic Cepheid Period-Luminosity relations from B to K bands, and a comparison 
  to LMC \emph{PL} relations}
  \titlerunning{Cepheids \emph{PL} relations}
  \authorrunning{P.~Fouqu\'e, P.~Arriagada, J.~Storm \emph{et al.}}
  \author{P.~Fouqu\'e\inst{1}, P.~Arriagada\inst{1,4} \and J.~Storm\inst{2} \and 
   T.G.~Barnes\inst{5,9} \and N.~Nardetto\inst{3} \and A.~M\'erand\inst{8} \and 
   P.~Kervella\inst{7} \and W.~Gieren\inst{6} \and D.~Bersier\inst{10} \and 
   G.F.~Benedict\inst{5} \and B.E.~McArthur\inst{5}}
  \institute{
    Observatoire Midi-Pyr\'{e}n\'{e}es, Laboratoire d'Astrophysique, UMR~5572, Universit\'{e} Paul Sabatier - Toulouse 3, 14 avenue Edouard Belin, 31400 Toulouse, France 
  \and Astrophysikalisches Institut Potsdam, An der Sternwarte 16, D-14482 Potsdam, Germany 
  \and Max-Planck-Institut f\"ur Radioastronomie, Infrared Interferometry Group, Auf dem H\"ugel 69, D-53121 Bonn, Germany 
  \and Departamento de Astronom\'{\i}a y Astrof\'{\i}sica, Pontificia Universidad Cat\'olica de Chile, Campus San Joaqu\'{\i}n, Vicu\~na Mackenna 4860, Casilla 306, Santiago 22, Chile. 
  \and University of Texas at Austin, McDonald Observatory, 1 University Station, C1402, Austin, TX 78712-0259, USA 
  \and Universidad de Concepci\'{o}n, Departamento de F\'{\i}sica, Grupo de Astronom\'{\i}a, Casilla 160-C, Concepci\'{o}n, Chile 
  \and Observatoire de Paris-Meudon, LESIA, UMR~8109, 5 place Jules Janssen, 92195 Meudon Cedex, France 
  \and Center for High Angular Resolution Astronomy, Georgia State University, PO Box 3965, Atlanta, Georgia 30302-3965, USA 
  \and currently on assignment to the National Science Foundation, 4201 Wilson Boulevard, Arlington, VA 22230, USA 
  \and Astrophysics Research Institute, Liverpool John Moores University, Twelve Quays House, Egerton Wharf, Birkenhead, CH41 1LD, UK 
  }
  \date{ Received / Accepted }
  \abstract
      {The universality of the Cepheid Period-Luminosity relations has been under discussion since metallicity effects have been
assumed to play a role in the value of the intercept and, more recently, of the slope of these relations.}
      {The goal of the present study is to calibrate the Galactic \emph{PL} relations in various photometric bands (from B to K) and
to compare the results to the well-established \emph{PL} relations in the LMC.}
      {We use a set of 59 calibrating stars, the distances of which are measured using five different distance indicators: Hubble
Space Telescope and revised Hipparcos parallaxes, infrared surface brightness and interferometric Baade-Wesselink parallaxes, and 
classical Zero-Age-Main-Sequence-fitting parallaxes for Cepheids belonging to open clusters or OB stars associations. A detailed 
discussion of absorption corrections and projection factor to be used is given.}
      {We find no significant difference in the slopes of the \emph{PL} relations between LMC and our Galaxy.}
      {We conclude that the Cepheid \emph{PL} relations have universal slopes in all photometric bands, not depending on 
the galaxy under study (at least for LMC and Milky Way). The possible zero-point variation with metal content is not discussed in the 
present work, but an upper limit of 18.50 for the LMC distance modulus can be deduced from our data.}
\maketitle

\section{Introduction} \label{sec:intro}

Soon after the release of Hipparcos measurements of Cepheid parallaxes 
\citep{Perryman1997}, \citet{Feast1997} intended to re-calibrate the Period-Luminosity 
relation of Cepheids using these parallaxes, with a conflictual result on the distance
to the Large Magellanic Cloud. However, the accuracy of the zero-point was quite limited
for such distant stars, and the derived LMC modulus depended heavily on the adopted PL slope.

Later, \citet{Benedict2002} obtained a parallax for $\delta$ Cep, using the Fine Guidance
Sensor~3 instrument on board HST, with an accuracy of 4\%, already better than the most
accurate Cepheid parallax from Hipparcos ($\alpha$ UMi, 6\%). Recently, \citet{Benedict2007} managed
to measure 9 additional Cepheid parallaxes, using FGS~1r on HST, with a mean accuracy of
8\%.

Finally, \citet{vanLeeuwen2007} published a revision of Hipparcos parallaxes, with a typical
improvement in accuracy for Cepheids of a factor two. $\alpha$ UMi is back at the top place, with
an accuracy of 1.6\%. But the second star, $\delta$ Cep, still has an accuracy slightly worse
than the HST measurement (5.2\% vs 4.1\%).

In parallel, Cepheid distances estimated via several variants of the Baade-Wesselink method 
have been secured. We use here two of these variants, the Infra-Red Surface Brightness
technique (hereafter, IRSB) \citep[e.g.:][]{Fouque1997} and the Interferometric 
Baade-Wesselink method (IBW), where Cepheid pulsation is directly measured with a 
long-baseline interferometer \citep[see, e.g.:][]{Kervella2004a}. The random uncertainty of 
distances obtained via these techniques is very small ($\sim 3\%$), but systematic 
uncertainties hamper a reliable calibration from this method alone. This mainly
comes from uncertainty in the so-called projection factor, which converts observed radial 
velocities to pulsation velocities. A recent discussion about which factor should be applied 
to these two techniques (IBW and IRSB) can be found in \citet{Nardetto2004}. This study, based 
on a hydrodynamical model of $\delta$~Cep has been then confirmed observationally by 
\citet{Merand2005} using the accurate distance measurement of this star by the HST.

Finally, the standard technique of deriving Cepheid distances from their association with an
open cluster is still valid \citep[see, e.g.:][]{Tammann2003} and will also be discussed here.

Our goal is to calibrate the \emph{PL} relation in various photometric bands (from B to K) using
Galactic Cepheids of known distances, and to compare the resulting relations to those in the
Large Magellanic Cloud (LMC).

\section{Observables} \label{sec:data}

\subsection{Photometry} \label{sec:phm}

For a Cepheid to enter our sample, we request photometry at least in B, V, $\Ic$, J and K bands.
We use the catalogue from \citet{Berdnikov2000} for the visible photometry, supplemented in
a few cases by data from \citet{Sandage2004} and \citet{Groenewegen1999}. These are 
intensity-mean values.

For the near-infrared bands, we started from the \citet{Berdnikov1996a} catalogue. However, it
does not include more recent measurements \citep{Barnes1997}, and it is not in a well-defined
system, although it claims to be in the CIT system. 


We therefore decided to use original intensity-mean values from \citet{Welch1984},
\citet{Laney1992}, \citet{Barnes1997} and to convert them to the 2MASS
system \citep{Skrutskie2006}, which is the only near-infrared system covering the whole sky. 
Transformations from CIT and SAAO systems to 2MASS use the updated conversion relations
from 2MASS web site\footnote{www.ipac.caltech.edu/2mass/releases/allsky/doc/sec6\_4b.html},
which replace the original Carpenter's  equations \citep{Carpenter2001}. When several sources are 
available, we take a weighted average of transformed intensity-mean values, using the number 
of measurements as a weight. When only \citet{Berdnikov1996a} values are available, we did not
go back to the original publication and assume that they are in the CIT system to convert them 
accordingly. If the intensity-mean values were not available from the original source 
(case of \citet{Barnes1997} and Y~Sgr in \citet{Welch1984}), we used the values as published 
in \citet{Groenewegen1999} (SAAO system) and the corresponding conversion. All these recipes 
obviously limit the accuracy of the resulting magnitudes, but we have to live with it, as measuring 
good near-infrared light curves of bright stars is difficult nowadays.

\subsection{Absorption} \label{sec:abs}

\subsubsection{Extinction} \label{sec:ext}

To correct observed magnitudes in each band for galactic absorption, we need an estimate of
the extinction E(B-V) and a reddening law. \citet{Tammann2003} (hereafter T03) discuss these issues in details:
they first adopt a revised Fernie's system \citep{Fernie1995} (hereafter F95) by converting extinction values from
various authors in Fernie's David Dunlap Observatory database of Galactic Classical 
Cepheids\footnote{www.astro.utoronto.ca/DDO/research/cepheids/ \\ table\_colourexcess.html}
to Fernie's own measurements system, before averaging them. Then, they test for a possible scale
error of this mean system and indeed detect that the mean F95 extinctions are too large by $\sim 5\%$.

In the present work, we make use of the recently published \citet{Laney2007} (hereafter L07) extinction values, 
based on B V $\Ic$ photometry. A test of these extinctions (solar metallicity assumed for all stars) vs. Fernie's 
original system (columns FE1 and FE2 in his database, FE1 if both present) for 155 stars in common reveals that 
the L07 values are precisely on the corrected T03 system (see their Eq.~2 and \Tab{tab:f95l07}).

We therefore adopt the L07 system as reference, and convert available data of each reference in Fernie's database 
using relations in \Tab{tab:f95l07}, where the symbols correspond to the column headings in the database. We replace
values from \citet{Bersier1996} by the revised and more extensive data set published in \citet{Bersier2002}. Note that 
we adopt only a scale factor when the intercept of the linear relation is not larger than its uncertainty. Also note
that we use the direct least-squares relation for conversion, which is the one giving unbiased results. This
explains why all the slopes are smaller than 1.

\begin{table*}
 \begin{center}
    \caption{Adopted conversion relations for extinction values, transforming $\EBV \ \mathrm{(X)}$ to
    the \citet{Laney2007} system, by $\EBV \ \mathrm{(L07)} = a \ \EBV \ \mathrm{(X)} + b$.}
    \begin{tabular}{llcccrc}
      \hline\hline \\
      Authors & F95 id & slope & intercept & $\sigma$ & N & weight \\
      \hline \\
      \citet{Parsons1975}       & PB   & $0.908 \pm 0.026$ &                    & 0.064 & 105 & 0.3 \\
      \citet{Yakimova1975}      & YT   & $0.863 \pm 0.017$ & $0.036 \pm 0.007$  & 0.034 & 82  & 1.2 \\
      \citet{Janot-Pacheco1976} & JP   & $0.873 \pm 0.077$ &                    & 0.089 & 30  & 0.1 \\
      \citet{Dean1978}          & DWC  & $0.961 \pm 0.012$ & $-0.014 \pm 0.005$ & 0.025 & 77  & 3.2 \\
      \citet{Pel1978}           & PE   & $0.945 \pm 0.029$ &                    & 0.047 & 64  & 0.5 \\
      \citet{Kron1979}          & KR   & $0.867 \pm 0.028$ &                    & 0.072 & 124 & 0.2 \\
      \citet{Feltz1980}         & FM   & $0.890 \pm 0.022$ & $0.039 \pm 0.013$  & 0.046 & 54  & 0.6 \\
      \citet{Dean1981}          & DE   & $0.857 \pm 0.034$ &                    & 0.038 & 35  & 0.9 \\
      \citet{Turner1987}        & TLE  & $0.938 \pm 0.050$ &                    & 0.037 & 21  & 1.0 \\
      \citet{Schechter1992}     & SACK & $0.861 \pm 0.073$ & $0.039 \pm 0.037$  & 0.034 & 9   & 1.2 \\
      \citet{Laney1993}         & LS   & $0.985 \pm 0.008$ &                    & 0.015 & 45  & -   \\
      \citet{Fernie1995}        & FE   & $0.952 \pm 0.010$ &                    & 0.029 & 147 & 1.9 \\
      \citet{Eggen1996}         & EG2  & $0.754 \pm 0.045$ & $0.064 \pm 0.013$  & 0.043 & 34  & 0.7 \\
      \citet{Bersier2002}       &      & $0.883 \pm 0.021$ & $0.035 \pm 0.009$  & 0.040 & 64  & 0.8 \\
      \hline
    \end{tabular}
    \label{tab:f95l07}
 \end{center}
\end{table*}

We then adopt a weighted mean of the individual measurements corrected with the above formulae, with the 
weights computed as the square inverse of the typical uncertainty of a given source, derived from the 
above rms dispersions: we assume for instance that L07 and DWC equally contribute to the observed dispersion of 
their conversion relation, giving 0.018 uncertainty to each, corresponding to a weight of 3.2 each. We do not use 
the LS values in the mean, which probably correspond to older derivations of L07 values. For convenience, we list 
the results for the whole sample of 158 stars in \Tab{tab:extinc}.

GT Car, SU Cru and BG Cru do not appear in L07: for GT Car, we use its \citet{Schechter1992} value of extinction from 
Fernie's database, converted to L07 system using the corresponding equation in \Tab{tab:f95l07}; similarly for SU Cru
and BG Cru, we use a weighted mean of their converted values of extinction from Fernie's database and \citet{Bersier2002}, 
after rejection of a discrepant value from \citet{Fernie1995} for BG Cru.


\begin{table*}
 \begin{center}
    \caption{Adopted weighted mean extinction values for the 155 Cepheids in \citet{Laney2007}, GT Car, SU Cru and BG Cru.}
    \begin{tabular}{lccr|lccr|lccr}
      \hline\hline
           &          &      &    &      &          &      &    &      &          &      &   \\
      Star & $E(B-V)$ & m.e. & N  & Star & $E(B-V)$ & m.e. & N  & Star & $E(B-V)$ & m.e. & N \\
      \hline
             &       &       &    &              &       &       &    &              &       &       &    \\
U Aql        & 0.360 & 0.010 & 10 & X Cyg        & 0.228 & 0.012 &  9 & X Pup        & 0.402 & 0.009 & 10 \\
SZ Aql       & 0.537 & 0.017 &  9 & SU Cyg       & 0.098 & 0.014 &  4 & RS Pup       & 0.457 & 0.009 & 10 \\
TT Aql       & 0.438 & 0.011 &  8 & SZ Cyg       & 0.571 & 0.015 &  5 & VZ Pup       & 0.459 & 0.011 &  7 \\
FF Aql       & 0.196 & 0.010 &  8 & TX Cyg       & 1.130 & 0.015 &  5 & WX Pup       & 0.301 & 0.015 &  4 \\
FM Aql       & 0.589 & 0.012 &  9 & VY Cyg       & 0.606 & 0.019 &  3 & WZ Pup       & 0.227 & 0.016 &  4 \\
FN Aql       & 0.483 & 0.010 &  7 & VZ Cyg       & 0.266 & 0.011 &  6 & AQ Pup       & 0.518 & 0.010 &  8 \\
V496 Aql     & 0.397 & 0.010 &  7 & CD Cyg       & 0.493 & 0.015 &  5 & AT Pup       & 0.191 & 0.010 &  8 \\
V600 Aql     & 0.798 & 0.016 &  4 & DT Cyg       & 0.042 & 0.011 &  7 & BN Pup       & 0.416 & 0.018 &  3 \\
$\eta$ Aql   & 0.130 & 0.009 & 13 & MW Cyg       & 0.635 & 0.017 &  4 & LS Pup       & 0.461 & 0.015 &  4 \\
RT Aur       & 0.059 & 0.013 &  5 & V386 Cyg     & 0.841 & 0.017 &  4 & S Sge        & 0.100 & 0.010 &  9 \\
RX Aur       & 0.263 & 0.012 &  5 & V402 Cyg     & 0.391 & 0.025 &  2 & GY Sge       & 1.187 & 0.170 &  2 \\
RW Cam       & 0.633 & 0.016 &  4 & V459 Cyg     & 0.730 & 0.019 &  3 & U Sgr        & 0.403 & 0.009 & 12 \\
RX Cam       & 0.532 & 0.011 &  7 & V532 Cyg     & 0.494 & 0.015 &  5 & W Sgr        & 0.108 & 0.011 &  6 \\
RY CMa       & 0.239 & 0.010 &  7 & V924 Cyg     & 0.261 & 0.025 &  2 & X Sgr        & 0.237 & 0.015 &  9 \\
RZ CMa       & 0.443 & 0.016 &  4 & V1726 Cyg    & 0.339 & 0.058 &  2 & Y Sgr        & 0.191 & 0.010 &  8 \\
SS CMa       & 0.553 & 0.011 &  8 & $\beta$ Dor  & 0.052 & 0.010 &  9 & WZ Sgr       & 0.431 & 0.011 &  8 \\
TW CMa       & 0.329 & 0.016 &  4 & W Gem        & 0.255 & 0.010 &  8 & XX Sgr       & 0.521 & 0.017 &  4 \\
U Car        & 0.265 & 0.010 &  8 & RZ Gem       & 0.563 & 0.026 &  4 & YZ Sgr       & 0.281 & 0.010 &  7 \\
V Car        & 0.169 & 0.011 &  6 & AA Gem       & 0.309 & 0.017 &  4 & AP Sgr       & 0.178 & 0.010 &  7 \\
SX Car       & 0.318 & 0.015 &  4 & AD Gem       & 0.173 & 0.019 &  3 & BB Sgr       & 0.281 & 0.009 & 10 \\
UX Car       & 0.112 & 0.012 &  7 & DX Gem       & 0.430 & 0.015 &  4 & V350 Sgr     & 0.299 & 0.010 &  8 \\
VY Car       & 0.237 & 0.009 &  9 & $\zeta$ Gem  & 0.014 & 0.011 &  7 & RV Sco       & 0.349 & 0.010 &  9 \\
WZ Car       & 0.370 & 0.011 &  8 & V Lac        & 0.335 & 0.017 &  4 & RY Sco       & 0.718 & 0.018 &  6 \\
XX Car       & 0.347 & 0.012 &  6 & X Lac        & 0.336 & 0.011 &  7 & KQ Sco       & 0.869 & 0.021 &  5 \\
XY Car       & 0.411 & 0.014 &  5 & Y Lac        & 0.207 & 0.016 &  4 & V482 Sco     & 0.336 & 0.013 &  6 \\
XZ Car       & 0.365 & 0.010 &  8 & Z Lac        & 0.370 & 0.011 &  7 & V500 Sco     & 0.593 & 0.016 &  4 \\
YZ Car       & 0.381 & 0.012 &  6 & RR Lac       & 0.319 & 0.014 &  6 & Y Sct        & 0.757 & 0.012 &  7 \\
AQ Car       & 0.165 & 0.012 &  6 & BG Lac       & 0.300 & 0.016 &  4 & Z Sct        & 0.492 & 0.013 &  6 \\
CT Car       & 0.570 & 0.025 &  2 & GH Lup       & 0.335 & 0.018 &  3 & RU Sct       & 0.921 & 0.012 &  7 \\
FR Car       & 0.334 & 0.014 &  5 & T Mon        & 0.181 & 0.011 & 12 & SS Sct       & 0.325 & 0.010 &  8 \\
GI Car       & 0.200 & 0.011 &  6 & SV Mon       & 0.234 & 0.010 &  8 & CM Sct       & 0.721 & 0.019 &  3 \\
GT Car       & 0.866 & 0.029 &  1 & TX Mon       & 0.485 & 0.013 &  6 & EV Sct       & 0.655 & 0.013 &  8 \\
$\ell$ Car   & 0.147 & 0.013 &  8 & CS Mon       & 0.506 & 0.018 &  3 & V367 Sct     & 1.231 & 0.025 &  2 \\
RW Cas       & 0.380 & 0.019 &  5 & CV Mon       & 0.722 & 0.022 &  7 & BQ Ser       & 0.815 & 0.025 &  2 \\
SU Cas       & 0.259 & 0.010 &  9 & S Mus        & 0.212 & 0.017 &  7 & ST Tau       & 0.368 & 0.031 &  3 \\
SW Cas       & 0.467 & 0.019 &  3 & RT Mus       & 0.344 & 0.021 &  6 & SZ Tau       & 0.295 & 0.011 &  6 \\
SZ Cas       & 0.794 & 0.013 &  4 & UU Mus       & 0.399 & 0.015 &  4 & R TrA        & 0.142 & 0.010 &  8 \\
CF Cas       & 0.553 & 0.011 &  7 & S Nor        & 0.179 & 0.009 & 10 & S TrA        & 0.084 & 0.009 & 10 \\
DD Cas       & 0.450 & 0.017 &  4 & U Nor        & 0.862 & 0.024 &  8 & $\alpha$ UMi & 0.003 & 0.013 &  6 \\
DL Cas       & 0.488 & 0.010 &  9 & TW Nor       & 1.157 & 0.014 &  4 & T Vel        & 0.289 & 0.010 &  8 \\
FM Cas       & 0.325 & 0.017 &  4 & QZ Nor       & 0.253 & 0.016 &  4 & V Vel        & 0.186 & 0.019 &  6 \\
V Cen        & 0.292 & 0.012 &  9 & V340 Nor     & 0.321 & 0.018 &  3 & RY Vel       & 0.547 & 0.010 &  9 \\
VW Cen       & 0.428 & 0.024 &  6 & Y Oph        & 0.645 & 0.015 &  9 & RZ Vel       & 0.299 & 0.010 &  8 \\
XX Cen       & 0.266 & 0.011 &  6 & BF Oph       & 0.235 & 0.010 &  7 & SW Vel       & 0.344 & 0.010 &  9 \\
AZ Cen       & 0.168 & 0.010 &  8 & RS Ori       & 0.352 & 0.012 &  7 & SX Vel       & 0.263 & 0.012 &  5 \\
KN Cen       & 0.797 & 0.091 &  6 & GQ Ori       & 0.249 & 0.014 &  5 & AX Vel       & 0.224 & 0.012 &  5 \\
V339 Cen     & 0.413 & 0.014 &  5 & SV Per       & 0.408 & 0.019 &  5 & CS Vel       & 0.737 & 0.029 &  4 \\
CR Cep       & 0.709 & 0.017 &  4 & UY Per       & 0.873 & 0.011 &  6 & DR Vel       & 0.656 & 0.014 &  5 \\
$\delta$ Cep & 0.075 & 0.010 &  9 & VX Per       & 0.475 & 0.011 &  7 & S Vul        & 0.727 & 0.042 &  3 \\
S Cru        & 0.166 & 0.010 &  8 & VY Per       & 0.945 & 0.013 &  6 & T Vul        & 0.064 & 0.011 &  7 \\
SU Cru       & 0.942 & 0.096 &  5 & AS Per       & 0.674 & 0.019 &  3 & U Vul        & 0.603 & 0.011 &  6 \\
AG Cru       & 0.212 & 0.018 &  5 & AW Per       & 0.489 & 0.012 &  6 & X Vul        & 0.742 & 0.019 &  7 \\
BG Cru       & 0.132 & 0.023 &  4 &              &       &       &    & SV Vul       & 0.461 & 0.022 &  8 \\
      \hline
    \end{tabular}
    \label{tab:extinc}
 \end{center}
\end{table*}

\subsubsection{Reddening law} \label{sec:redlaw}

For the reddening law, we adopt the \citet{Cardelli1989} system, contrarily to our previous works,
where the reddening law was derived from \citet{Laney1993} (hereafter LS93), and \citet{Caldwell1987}. 
This is not a matter of preference, but we had to derive absorption ratios for the 2MASS system, and 
Cardelli's formulae were suitable for this purpose: we adopted isophotal wavelengths from \citet{Cohen2003}. 
For the Cousins bands not given in \citet{Cardelli1989}, we used isophotal wavelengths from 
\citet{Bessell1998}. We neglect the small difference with effective wavelengths suitable for 
Cepheids colours.

Cardelli's formulae depend on the total-to-selective absorption ratio in V band, $\Rv$. This ratio
slightly depends on the star colour and on extinction, but, as in \citet{Fouque2003}, we adopt a 
constant value, because the colour dependency is not well established and in any case small for the colour 
range of Cepheids. We adopt the mean value derived by \citet{Sandage2004}, namely $\Rv=3.23$, and
$\Rb=\Rv+1$. \Tab{tab:absrat} compares published values of these ratios in various bands.

\begin{table}
 \begin{center}
    \caption{Comparaison of published total-to-selective absorption ratios in various bands.}
    \begin{tabular}{lccc}
      \hline\hline \\
      Filter & $R (\lambda)$ & Reference \\
      \hline \\
      V     & 3.26 & \citet{Berdnikov1996b} \\
      V     & 3.30 & \citet{Fouque2003}     \\
      V     & 3.23 & \citet{Sandage2004}    \\
      V     & 3.1  & \citet{Benedict2007}   \\
      V     & 3.23 & this work              \\
      $\Ic$ & 1.86 & \citet{Berdnikov1996b} \\
      $\Ic$ & 1.99 & \citet{Fouque2003}     \\
      $\Ic$ & 1.95 & \citet{Sandage2004}    \\
      $\Ic$ & 1.73 & \citet{Benedict2007}   \\
      $\Ic$ & 1.96 & this work              \\
      J     & 0.82 & \citet{Fouque2003}     \\
      J     & 0.88 & \citet{Benedict2007}   \\
      J     & 0.94 & this work              \\
      H     & 0.48 & \citet{Fouque2003}     \\
      H     & 0.58 & this work              \\
      K     & 0.30 & \citet{Fouque2003}     \\
      K     & 0.34 & \citet{Benedict2007}   \\
      $\Ks$ & 0.38 & this work              \\
      \hline
    \end{tabular}
    \label{tab:absrat}
 \end{center}
\end{table}

The small isophotal wavelength differences between SAAO and 2MASS systems do not explain the difference
in infrared total-to-selective absorption ratios between \citet{Fouque2003}, adopted from LS93, and the
present work. The source of this discrepancy is the different approach to derive the reddening law used by  
\citet{Cardelli1989} and LS93.

\Tab{tab:redlaw} gives our adopted reddening law for the various photometric bands, and the LS93
values for the SAAO infrared bands.

\begin{table}
 \begin{center}
    \caption{Adopted reddening law from \citet{Cardelli1989}, using isophotal wavelengths 
     (in $\mu$m) from \citet{Bessell1998} for Cousins bands and from \citet{Cohen2003} for 2MASS bands, 
     and $\Rv=3.23$. For comparison, LS93 values in the SAAO infrared bands are also given.}
    \begin{tabular}{lccc}
      \hline\hline \\
      Filter & $\lambda_{\mathrm{iso}}$ & $A(\lambda)/A(V)$ & LS93 \\
      \hline \\
      B     & 0.438 & 1.310 &       \\
      V     & 0.545 & 1     &       \\
      $\Rc$ & 0.641 & 0.845 &       \\
      $\Ic$ & 0.798 & 0.608 &       \\
      J     & 1.235 & 0.292 & 0.249 \\
      H     & 1.662 & 0.181 & 0.147 \\
      $\Ks$ & 2.159 & 0.119 & 0.091 \\
      \hline
    \end{tabular}
    \label{tab:redlaw}
 \end{center}
\end{table}

\subsection{Parallaxes} \label{sec:dis}

We will now give some details about the five methods we use to measure
Cepheid parallaxes.

\subsubsection{Trigonometric parallaxes} \label{sec:hstdis}

We use two sources of trigonometric parallaxes: HST parallaxes, published by \citet{Benedict2002} 
and \citet{Benedict2007}, and the revised Hipparcos parallaxes of Cepheids, from \citet{vanLeeuwen2007}.
Table~1 of the latter reference compares the two sources for the 10 Cepheids measured by HST.
It is clear that even for this selected sample of the best revised Hipparcos parallaxes, the HST
measurements are slightly superior. The case of Y Sgr, where the discrepancy is much larger than
the combined uncertainties, is worrying.

In the present study, we adopt the HST parallaxes as reference, and make use of the revised Hipparcos parallaxes
only if their accuracy is better than 30\% and they are not common to the HST sample: there are 8 such stars,
5 of them being classified as first overtone pulsator.

\subsubsection{IRSB parallaxes} \label{sec:irsbdis}

The Infra-Red Surface Brightness technique can give very precise parallaxes when the radial
velocity and light curves are well defined. However, they directly depend on the adopted value of
the projection factor, which we will discuss in detail here.

For the visible surface brightness versus $\VK$ colour relation, we still use \citet{Fouque1997} 
for consistency with previous work, although it has been superseded by more recent studies. 
In fact, all the more recent calibrations (\citet{Nordgren2002}; \citet{Kervella2004c}; 
\citet{Groenewegen2004}), either based on giants or on Cepheids, confirm the validity of this calibration
and of the hypothesis that stable giant stars and pulsating Cepheids obey the same relation.

Since the publication of our last paper on IRSB distances in \citet{Barnes2005}, which contained
38 Cepheids, many new measurements have been added to our database to reach a total of 70 stars. 
A separate publication describing these new results and giving references to the spectroscopic and photometric
data involved in this work is in preparation \citep{Storm2008}.

The projection factor $p$ is defined as the ratio of the pulsational velocity to the radial velocity.
Early studies \citep[e.g.:][]{Carroll1928} only considered geometrical effects, namely limb darkening 
and atmospheric expansion at constant velocity, and arrived at values between 1.375 and 1.412 for visible 
linear limb-darkening coefficients $u_V$ between 0.8 and 0.6, respectively. This is well represented by Eq.~6 in 
\citet{Nardetto2006}, $p = 1.52 - 0.18 \  u_V$. As the limb darkening is mainly related to the effective 
temperature and the surface gravity, it varies from one Cepheid to another. We already see that a single 
$p$-factor for all Cepheids cannot be but an over-simplistic approximation, and even a single $p$-factor for 
a given Cepheid is an approximation, because its temperature and surface gravity vary along the pulsation cycle
(for details, see, e.g.: \citet{Marengo2002}, \citet{Marengo2003}).

And this is just the tip of the iceberg: $p$ also depends on the proper amplitude of the pulsation velocity, 
on the lines under study, on the way the radial velocities are measured, on the spectral resolution, as first 
shown by \citet{Parsons1972}. Following the work of \citet{Hindsley1986}, who adapted the value of $p$ for radial velocities measured 
through cross-correlation techniques, where many lines are mixed into the measurement of the correlation peak, 
we adopted as an approximation of all these effects $p = 1.39 - 0.03 \log P$ to represent the variation of the 
$p$-factor with the period of the Cepheid (hereafter \emph{p\,(P)} relation). This is obviously a rather crude 
approximation, but it is difficult to be more specific. In \citet{Gieren2005}, we revised the coefficients of 
this relation, because they lead to an apparent variation of the LMC distance with Cepheid period, in a sample 
of 13 LMC Cepheids for which an IRSB distance was measured. At the same time, the new relation 
$p = 1.58 - 0.15 \log P$ leads to a better agreement between LMC and Galactic slopes of the \emph{PL} relations 
in various photometric bands, raising again the hope of universal \emph{PL} relations. Improvements in the modelling 
investigations also permit a better understanding of the various components of the $p$-factor, but they lead 
to values more in agreement with our old relation \citep[see e.g.:][]{Nardetto2004}. We will test the most 
recent relation based on these models and high resolution spectroscopic observations, as published in 
\citet{Nardetto2007}, namely $p = 1.366 - 0.075 \log P$. In the case of suspected first overtone pulsators, 
we use the observed period to compute the corresponding $p$-factor, as in our previous studies, not the 
corrected period of fundamental pulsation.

\subsubsection{Interferometric Baade-Wesselink parallaxes} \label{sec:ibwdis}

The originality of the Interferometric Baade-Wesselink method compared to the classical IRSB technique is to 
measure directly the angular amplitude of the pulsation of the star, instead of deducing it from photometric colours. 
In the last few years, the considerable improvement of the available long-baseline interferometry (LBI)
instrumentation has shed a new light on the IBW technique. The first interferometric observations of a Cepheid were 
obtained by \citet{Mourard1997}, followed by observations from almost all operating interferometers 
(\citet{Lane2000}; \citet{Nordgren2000}; \citet{Kervella2001}; \citet{Kervella2004b}; \citet{Merand2005}; 
\citet{Merand2006}; \citet{Merand2007}).

The main difficulty in observing Cepheids by LBI is that they are rare and therefore relatively distant stars. As a 
consequence, they present small angular sizes, even for the closest ones. The 
Cepheid which has the largest angular diameter is $\ell$\,Car, with 3\,mas, and a baseline of approximately 180\,m is
already required to fully resolve it in the infrared. As the pulsation amplitude of a Cepheid is about 20\% of its 
size, the detection of the pulsation, necessary to apply the IBW method, is an even more challenging objective. On 
the 10 nearest Cepheids, an accurate measurement of the amplitude of the pulsation (to a few percent accuracy) 
requires a baseline length of 150 to 300\,m.

As a consequence, the distances to only eight Cepheids have been measured using the true IBW technique: 
$\delta$\,Cep \citep{Merand2005}, $\eta$\,Aql \citep{Lane2002}, $\beta$\,Dor (\citet{Kervella2004b}; \citet{Davis2006a}), 
$\zeta$\,Gem \citep{Lane2002}, Y\,Oph \citep{Merand2007}, $\ell$\,Car (\citet{Kervella2004a}; \citet{Davis2006b}),
W Sgr \citep{Kervella2004b}, and Y Sgr \citep{Merand2008}. We choose in the present paper to focus on these stars only.

Before integration, we interpolated the spectroscopic radial velocities using node-constrained cubic splines, as described by 
\citet{Merand2007}. In the literature, the hypotheses used for the application of the IBW method may substantially differ 
among authors. In order to get an homogeneous set, we recomputed the parallaxes using the original interferometric uniform disk 
diameters, a limb-darkening model from \citet{Claret2000}, and the $p$-factor adopted in \Sec{sec:adopdis}. 
The derived parallaxes are given in \Tab{tab:ibwpar}: three of them have large uncertainties and are therefore excluded from the 
following analysis. The recent discovery of circumstellar envelopes around most Cepheids for which interferometric measures exist 
(\citet{Kervella2006}, \citet{Merand2006}, \citet{Merand2007}) has a clear impact on this method, the importance of which must 
still be assessed.

\begin{table}
 \begin{center}
    \caption{Adopted parallaxes measured by the Interferometric Baade-Wesselink technique. The adopted $p$-factor is
    listed in the last column.}
    \begin{tabular}{llccc}
      \hline\hline
           &          &       &                &     \\
      Star & $\log P$ & $\pi$ & $\sigma (\pi)$ & $p$ \\
      \hline
             &          &       &       &       \\        
$\eta$ Aql   & 0.855930 & 3.31  & 0.05  & 1.302 \\
$\ell$ Car   & 1.550816 & 1.90  & 0.07  & 1.250 \\
$\delta$ Cep & 0.729678 & 3.52  & 0.10  & 1.311 \\
$\beta$ Dor  & 0.993131 & 3.05  & 0.98  & 1.292 \\
$\zeta$ Gem  & 1.006507 & 2.91  & 0.31  & 1.291 \\
Y Oph        & 1.233609 & 2.16  & 0.08  & 1.273 \\
W Sgr        & 0.880522 & 2.76  & 1.23  & 1.300 \\
Y Sgr        & 0.761428 & 1.96  & 0.62  & 1.309 \\
      \hline
    \end{tabular}
    \label{tab:ibwpar}
 \end{center}
\end{table}

\subsubsection{Open cluster parallaxes} \label{sec:clusdis}

We adopt the parallaxes for Cepheids belonging to open clusters or associations from the recent compilation
by \citet{Turner2002}. No correction for the underlying assumed Pleiades distance modulus (5.56) has been attempted, 
because the validity of such a correction is questionable \citep{Feast1999}. The geometrical parallax of RS~Pup is adopted
from \citet{Sandage2004}.

The basic hypothesis underlying this kind of parallax measurement is that the Cepheid is indeed
associated with the cluster. We will see below in \Sec{sec:adopdis} which Cepheids are dubious members of their
association or cluster.

Parallax uncertainties as published by \citet{Turner2002} seem too small, sometimes reaching 1\%. Based on the more
realistic uncertainties published by \citet{Hoyle2003}, we have set a minimum accuracy of 5\%.

\subsubsection{Adopted parallaxes} \label{sec:adopdis}

The status of known Cepheid parallaxes is the following, without taking into account Hipparcos parallaxes with a large
uncertainty: 81 stars have a parallax from at least one distance indicator, 22 from two, 10 from three, 4 from four, 
and only 1 from the five distance indicators: see \Tab{tab:meanpar}. 
Among the 10 stars with HST parallaxes, all have an Hipparcos parallax, 7 have IRSB, 6 have IBW and 2 have ZAMS measurements, 
but all these measurements largely vary in quality. Moreover, we have to decide about the choice of the final \emph{p\,(P)} 
relation to adopt. Obviously, the choice of the slope of the \emph{p\,(P)} relation with $\log P$ has an influence on the
slope of the \emph{PL} relations, in the sense that a shallower slope of this relation (such as the small dependence of 
the classical relation) leads to a steeper slope in the \emph{PL} relations. On the other hand, there are not enough HST 
parallaxes to settle this slope definitively (there is only one long-period star in the HST sample).

In \citet{Gieren2005}, the slope of the \emph{p\,(P)} relation with $\log P$ was constrained by individual distance
measurements of 13 LMC Cepheids, so that the LMC distance modulus derived from each of these stars did not depend on
$\log P$. Although this is a reasonable constraint, a precise determination was hampered by several facts: the period
distribution of the Cepheids was not optimal, with all short-period Cepheids belonging to one single cluster, namely
NGC~1866; the corrections for the position of each Cepheid with respect to the geometry of the LMC were model-dependent; 
the uncertainties of the long-period Cepheids distance moduli were too small to explain the observed
dispersion of the derived LMC distance moduli among these stars. A revision of this determination with additional stars 
is in progress. For the time being, we therefore prefer to rely on the slope of the model relation from \citet{Nardetto2007}.

However, the zero-point of this relation is only valid for a well-defined way of measuring radial velocities from the
observed metallic lines, namely the first moment of the line. As most of our radial velocities come from 
cross-correlation measurements, where the peak of the cross-correlation is generally measured by a gaussian fit, there is
no reason that the model zero-point applies exactly to these measurements. As we want to use the HST parallaxes as the 
absolute system, we prefer to determine a preliminary \emph{PL} relation based on IRSB parallaxes, and verify that the
zero-point of the \emph{p\,(P)} relation does not lead to a significant shift of the 10 HST Cepheids when adopting their HST
parallaxes. For this purpose, we only use the $\Wvi$ (see \Sec{sec:gplrel} for definition) and the $\Ks$ \emph{PL} relations, 
as the less dispersed relations being representative of the optical and infrared bands. 

We first need to define a clean sample of IRSB calibrators, among the 70 stars with available IRSB parallaxes. After rejection of
5 known first overtone pulsators, of 2 stars with poor data fit, of 3 stars with poor sampling of the K light curve, of 4
additional outliers, and of 2 long-period Cepheids with variable periods, we are left with 54 calibrators. The mean residual of 
the 10 stars with HST parallaxes from the two \emph{PL} relations is $-0.01 \pm 0.03$ in $\Wvi$ and $-0.05 \pm 0.03$ in $\Ks$. 
They are not significant, and we therefore also adopt the zero-point of the model \emph{p\,(P)} relation from \citet{Nardetto2007}.

\begin{figure*}
  \begin{center}
    \includegraphics[width=14cm]{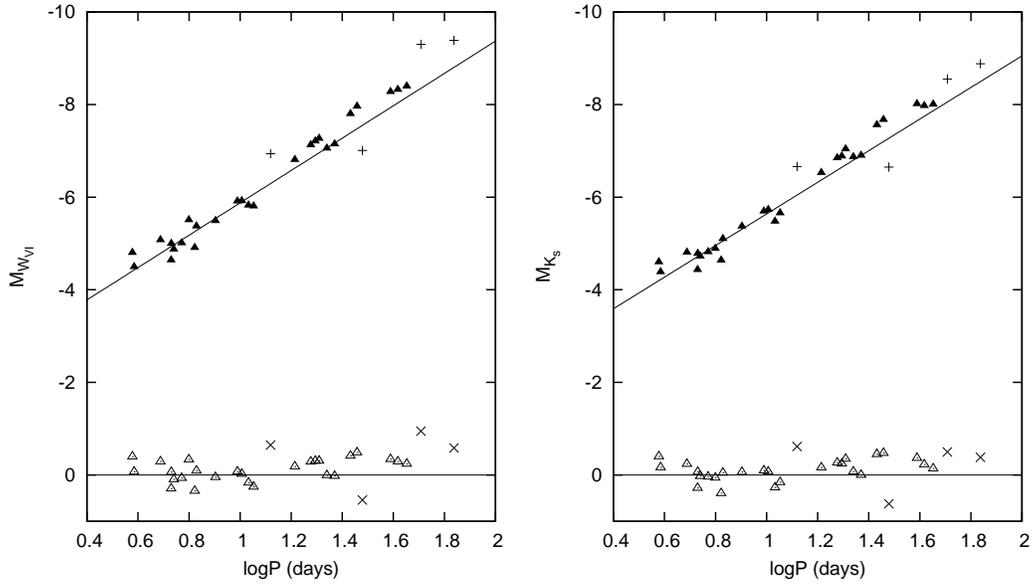}
    \caption{Comparison of absolute magnitudes derived from ZAMS parallaxes (triangles) for 30 stars with the \emph{PL} relations based 
    on 54 IRSB parallaxes in $\Wvi$ and $\Ks$ bands (solid lines). The bottom panel shows residuals with respect to the IRSB \emph{PL} 
    relations. Long-period Cepheids appear to be progressively shifted. Rejected stars (see text) are marked with pluses and crosses.}
    \label{fig:zams}
  \end{center}
\end{figure*}

If we compare the absolute magnitudes in $\Wvi$ and $\Ks$ of the Cepheids using their ZAMS parallaxes to the preliminary \emph{PL} 
relations based on IRSB parallaxes, we find in a sample of 26 Cepheids\footnote{We have rejected as dubious ZAMS parallaxes of GT Car 
and AQ Pup, and long-period stars GY Sge and S Vul} a mean shift of $-0.09 \pm 0.04$ mag, corresponding to ZAMS parallaxes being too 
small by 4\%. Part of the explanation of 
this shift may lie in the assumed Pleiades distance modulus in \citet{Turner2002}, which is 5.56, while the most accurate value given 
by \citet{Soderblom2005} is $5.63 \pm 0.02$, a difference of 3\%. However, a tendency to get more discrepant parallaxes for long-period
Cepheids, as shown in \Fig{fig:zams} which displays all 30 stars, leads us not to use the ZAMS parallaxes in the present study. 
A possible explanation of this effect is that long-period Cepheids are generally assigned to OB associations, where the membership is 
more difficult to establish. The small shift of ZAMS parallaxes is the most probable explanation to the too large zero-point of the 
\citet{Gieren2005} \emph{p\,(P)} relation (1.58), which was adjusted to make IRSB distances correspond to ZAMS ones.

It is interesting to compare the dispersion of the various indicators with respect to the preliminary \emph{PL} relations
based on 54 IRSB parallaxes, for which we have 0.15 in $\Ks$ and 0.17 in $\Wvi$: for the HST, 
the rms dispersion is 0.10 both in $\Ks$ and $\Wvi$; for the IBW, it is 0.18 in $\Ks$ and 0.21 in $\Wvi$, after
rejection of Y~Oph; and for the ZAMS, it amounts to 0.22 in $\Ks$ and 0.23 in $\Wvi$, after rejection of GT~Car and AQ~Pup. 
Clearly, the HST parallaxes are superior in accuracy to other distance indicators.

We decided not to average parallaxes for a given star among the different techniques, but to use HST parallaxes when 
available (to be consistent with our adjusted zero-point based on HST system) and IRSB, IBW or revised Hipparcos parallaxes 
if not. ZAMS parallaxes are not used due to their slight offset. The catalogue of adopted parallaxes is given in 
\Tab{tab:meanpar}. Suspected first overtone Cepheids are marked as 'FO' and will not be used in the following calibrations.

\begin{table*}
 \begin{center}
    \caption{Adopted parallax of our calibrators. Five different distance indicators
     have been considered, but only one has been chosen: HST (1), IBW (2), IRSB (3), ZAMS (4), HIP (5). 
     Last column lists the available indicators, the first one being the adopted one. When a distance
     indicator is of low quality or leads to rejection, it is given in brackets.
     FO marks the stars suspected to pulse in the first overtone mode. DM means double mode pulsator.}
    \begin{tabular}{llcclc|llcclc}
      \hline\hline
           &          &       &                &        &      &      &          &       &                &        &      \\
      Star & $\log P$ & $\pi$ & $\sigma (\pi)$ & source & mode & Star & $\log P$ & $\pi$ & $\sigma (\pi)$ & source & mode \\
      \hline
             &          &       &       &        &    &              &          &       &       &       &    \\
U Aql        & 0.846591 & 1.65  & 0.08  & 3      &    & S Nor        & 0.989194 & 1.06  & 0.05  & 34    &    \\
SZ Aql       & 1.234029 & 0.46  & 0.01  & 3      &    & U Nor        & 1.101875 & 0.81  & 0.04  & 3     &    \\
TT Aql       & 1.138459 & 0.99  & 0.03  & 3      &    & TW Nor       & 1.0329   & 0.52  & 0.03  & 4     &    \\
FF Aql       & 0.650397 & 2.81  & 0.18  & 15(3)  &    & QZ Nor       & 0.578244 & 0.79  & 0.10  & 34    &    \\
FM Aql       & 0.786342 & 0.85  & 0.05  & (3)    &    & V340 Nor     & 1.052579 & 0.56  & 0.11  & 34    &    \\
FN Aql       & 0.976878 & 0.90  & 0.03  & (3)    &    & Y Oph        & 1.233609 & 1.81  & 0.13  & 32    &    \\
V496 Aql     & 0.832958 & 1.07  & 0.11  & 3      &    & BF Oph       & 0.609329 & 1.57  & 0.08  & (3)   &    \\
$\eta$ Aql   & 0.855930 & 4.15  & 0.24  & 325    &    & X Pup        & 1.4143   & 0.47  & 0.03  & (3)   &    \\
RT Aur       & 0.571489 & 2.40  & 0.19  & 1(35)  &    & RS Pup       & 1.617420 & 0.55  & 0.03  & 34    &    \\
U Car        & 1.588970 & 0.67  & 0.03  & 34     &    & VZ Pup       & 1.364945 & 0.27  & 0.02  & (3)   &    \\
V Car        & 0.82586  & 0.96  & 0.14  & 3      &    & AQ Pup       & 1.478624 & 0.33  & 0.02  & 3(4)  &    \\
VY Car       & 1.276818 & 0.55  & 0.03  & 34     &    & BN Pup       & 1.135867 & 0.26  & 0.02  & 3     &    \\
WZ Car       & 1.361977 & 0.29  & 0.01  & 3      &    & LS Pup       & 1.150646 & 0.21  & 0.02  & 3     &    \\
GT Car       & 1.119    & 0.10  & 0.02  & (4)    &    & S Sge        & 0.923352 & 1.48  & 0.05  & 3     &    \\
$\ell$ Car   & 1.550816 & 2.01  & 0.20  & 1235   &    & GY Sge       & 1.7081   & 0.32  & 0.01  & (34)  &    \\
SU Cas       & 0.289884 & 2.25  & 0.17  & 354    & FO & U Sgr        & 0.828997 & 1.77  & 0.06  & 34    &    \\
CF Cas       & 0.6880   & 0.29  & 0.02  & 4      &    & W Sgr        & 0.880522 & 2.28  & 0.20  & 15(2) &    \\
DL Cas       & 0.9031   & 0.60  & 0.03  & 4      &    & X Sgr        & 0.845907 & 3.00  & 0.18  & 15    &    \\
V Cen        & 0.739882 & 1.65  & 0.12  & 34     &    & Y Sgr        & 0.761428 & 2.13  & 0.29  & 13(25)&    \\
VW Cen       & 1.177138 & 0.28  & 0.01  & 3      &    & WZ Sgr       & 1.339443 & 0.57  & 0.02  & 34    &    \\
XX Cen       & 1.039548 & 0.66  & 0.04  & 3      &    & YZ Sgr       & 0.980171 & 1.00  & 0.12  & (3)   &    \\
KN Cen       & 1.531857 & 0.27  & 0.02  & 3      &    & BB Sgr       & 0.821971 & 1.27  & 0.05  & 34    &    \\
$\delta$ Cep & 0.729678 & 3.66  & 0.15  & 1235(4)&    & V350 Sgr     & 0.712165 & 1.07  & 0.06  & 3     &    \\
SU Cru       & 1.1088   & 0.62  & 0.08  & 3      &    & RY Sco       & 1.307927 & 0.85  & 0.04  & 3     &    \\
BG Cru       & 0.524    & 2.23  & 0.30  & 5      & FO & KQ Sco       & 1.4577   & 0.35  & 0.04  & 4     &    \\
X Cyg        & 1.214482 & 0.86  & 0.02  & 34     &    & RU Sct       & 1.29448  & 0.58  & 0.05  & 34    &    \\
SU Cyg       & 0.584952 & 1.19  & 0.05  & 34     &    & SS Sct       & 0.564814 & 1.32  & 0.15  & (3)   &    \\
VZ Cyg       & 0.687034 & 0.54  & 0.02  & 3      &    & EV Sct       & 0.490098 & 0.60  & 0.12  & (3)4  & FO \\
CD Cyg       & 1.232334 & 0.39  & 0.01  & 3      &    & V367 Sct     & 0.7989   & 0.61  & 0.03  & 4     & DM \\
DT Cyg       & 0.397804 & 2.19  & 0.33  & 5(3)   & FO & SZ Tau       & 0.498166 & 1.95  & 0.11  & 345   & FO \\
$\beta$ Dor  & 0.993131 & 3.14  & 0.16  & 135(2) &    & $\alpha$ UMi & 0.5990   & 7.72  & 0.12  & 5     & FO \\
$\zeta$ Gem  & 1.006507 & 2.78  & 0.18  & 1245   &    & T Vel        & 0.666501 & 0.99  & 0.01  & 3     &    \\
X Lac        & 0.735997 & 0.37  & 0.05  & 3      & FO & RY Vel       & 1.449158 & 0.45  & 0.03  & 3     &    \\
Y Lac        & 0.635863 & 0.44  & 0.02  & 3      &    & RZ Vel       & 1.309564 & 0.70  & 0.03  & 345   &    \\
Z Lac        & 1.036854 & 0.53  & 0.01  & 3      &    & SW Vel       & 1.370016 & 0.42  & 0.01  & 34    &    \\
BG Lac       & 0.726883 & 0.59  & 0.03  & 3      &    & CS Vel       & 0.771201 & 0.33  & 0.03  & 34    &    \\
GH Lup       & 0.967448 & 0.89  & 0.15  & 3      &    & S Vul        & 1.837426 & 0.24  & 0.09  & (34)  &    \\
T Mon        & 1.431915 & 0.72  & 0.03  & 34     &    & T Vul        & 0.646934 & 1.90  & 0.23  & 135   &    \\
CV Mon       & 0.730685 & 0.63  & 0.05  & 34     &    & U Vul        & 0.902584 & 1.46  & 0.06  & 3     &    \\
S Mus        & 0.98498  & 1.22  & 0.08  & 35     &    & SV Vul       & 1.652569 & 0.46  & 0.01  & 34    &    \\
UU Mus       & 1.065819 & 0.33  & 0.04  & 3      &    &              &          &       &       &       &    \\
      \hline
    \end{tabular}
    \label{tab:meanpar}
 \end{center}
\end{table*}

\section{Galactic Period-Luminosity relations} \label{sec:gplrel}

From the adopted intensity-mean values in the various photometric bands, the extinction value from \Tab{tab:extinc}, 
the adopted reddening law from \Tab{tab:redlaw}, and the adopted parallax from \Tab{tab:meanpar}, we derive 
the absolute magnitudes in B, V, $\Rc$, $\Ic$, J, H and $\Ks$ of our 59 calibrators, listed in \Tab{tab:absmag}. 
As the uncertainty of these absolute magnitudes depend not only on distance uncertainties, but also on extinction and 
reddening law uncertainties, we choose not to weigh the derived \emph{PL} relations.

\begin{table*}
 \begin{center}
    \caption{Adopted absolute magnitudes of the 59 calibrators in 7 photometric bands (from B to $\Ks$) and for two 
             Wesenheit indices, $\Wvi$ and $\Wbi$.}
    \begin{tabular}{llccccccccc}
      \hline\hline
           &          &       &       &        &        &         &         &       &       &        \\
      Star & $\log P$ & $\Mb$ & $\Mv$ & $\Mrc$ & $\Mic$ & $\Mwvi$ & $\Mwbi$ & $\Mj$ & $\Mh$ & $\Mks$ \\
      \hline
             &          &       &       &       &       &       &       &       &       &       \\
RT Aur       & 0.571489 & -2.31 & -2.84 &       & -3.40 & -4.28 & -4.35 & -3.94 & -4.15 & -4.24 \\
QZ Nor       & 0.578244 & -1.83 & -2.47 & -2.85 & -3.15 & -4.21 & -4.29 & -3.67 & -3.91 & -4.01 \\
SU Cyg       & 0.584952 & -2.61 & -3.08 & -3.38 & -3.63 & -4.47 & -4.50 & -4.08 & -4.28 & -4.36 \\
Y Lac        & 0.635863 & -2.79 & -3.31 & -3.65 & -3.90 & -4.81 & -4.86 & -4.33 & -4.58 & -4.66 \\
T Vul        & 0.646934 & -2.48 & -3.06 & -3.41 & -3.66 & -4.58 & -4.68 & -4.12 & -4.36 & -4.45 \\
FF Aql       & 0.650397 & -2.46 & -3.02 & -3.35 & -3.64 & -4.60 & -4.65 & -4.09 & -4.29 & -4.37 \\
T Vel        & 0.666501 & -2.28 & -2.93 & -3.32 & -3.64 & -4.74 & -4.81 & -4.13 & -4.41 & -4.51 \\
VZ Cyg       & 0.687034 & -2.62 & -3.23 & -3.62 & -3.88 & -4.90 & -4.98 & -4.36 & -4.60 & -4.70 \\
V350 Sgr     & 0.712165 & -2.74 & -3.34 & -3.73 & -4.02 & -5.06 & -5.13 & -4.51 & -4.78 & -4.85 \\
BG Lac       & 0.726883 & -2.56 & -3.22 & -3.62 & -3.91 & -4.99 & -5.08 & -4.36 & -4.64 & -4.73 \\
$\delta$ Cep & 0.729678 & -2.88 & -3.47 & -3.87 & -4.11 & -5.11 & -5.18 & -4.55 & -4.82 & -4.91 \\
CV Mon       & 0.730685 & -2.46 & -3.04 & -3.51 & -3.78 & -4.94 & -4.93 & -4.35 & -4.62 & -4.72 \\
V Cen        & 0.739882 & -2.45 & -3.03 & -3.40 & -3.68 & -4.68 & -4.74 & -4.16 & -4.43 & -4.52 \\
Y Sgr        & 0.761428 & -2.57 & -3.23 & -3.63 & -3.95 & -5.07 & -5.15 & -4.45 & -4.75 & -4.83 \\
CS Vel       & 0.771201 & -2.48 & -3.09 & -3.54 & -3.79 & -4.88 & -4.93 & -4.33 & -4.59 & -4.69 \\
BB Sgr       & 0.821971 & -2.74 & -3.45 & -3.88 & -4.19 & -5.35 & -5.45 & -4.70 & -4.99 & -5.08 \\
V Car        & 0.82586  & -2.58 & -3.29 &       & -4.01 & -5.13 & -5.24 & -4.50 & -4.79 & -4.89 \\
U Sgr        & 0.828997 & -2.67 & -3.36 & -3.79 & -4.10 & -5.24 & -5.34 & -4.61 & -4.89 & -4.97 \\
V496 Aql     & 0.832958 & -2.63 & -3.39 & -3.85 & -4.16 & -5.34 & -5.48 & -4.67 & -4.95 & -5.02 \\
X Sgr        & 0.845907 & -3.31 & -3.82 & -4.17 & -4.43 & -5.38 & -5.40 & -4.87 & -5.10 & -5.17 \\
U Aql        & 0.846591 & -2.97 & -3.65 & -4.07 & -4.35 & -5.43 & -5.55 & -4.87 & -5.12 & -5.21 \\
$\eta$ Aql   & 0.855930 & -2.77 & -3.43 &       & -4.14 & -5.23 & -5.32 & -4.63 & -4.91 & -5.00 \\
W Sgr        & 0.880522 & -3.25 & -3.89 & -4.31 & -4.58 & -5.64 & -5.72 & -5.10 &       & -5.47 \\
U Vul        & 0.902584 & -3.32 & -3.99 & -4.47 & -4.75 & -5.92 & -5.99 & -5.17 & -5.41 & -5.46 \\
S Sge        & 0.923352 & -3.16 & -3.86 & -4.27 & -4.57 & -5.67 & -5.79 & -5.07 & -5.35 & -5.44 \\
GH Lup       & 0.967448 & -2.83 & -3.71 & -4.21 & -4.56 & -5.88 & -6.07 & -5.14 & -5.47 & -5.59 \\
S Mus        & 0.98498  & -3.50 & -4.13 & -4.50 & -4.80 & -5.85 & -5.93 & -5.27 & -5.55 & -5.66 \\
S Nor        & 0.989194 & -3.25 & -4.02 & -4.46 & -4.80 & -6.01 & -6.14 & -5.36 & -5.68 & -5.79 \\
$\beta$ Dor  & 0.993131 & -3.18 & -3.93 & -4.34 & -4.68 & -5.84 & -5.97 & -5.17 & -5.49 & -5.59 \\
$\zeta$ Gem  & 1.006507 & -3.13 & -3.93 &       & -4.70 & -5.90 & -6.06 & -5.31 &       & -5.71 \\
Z Lac        & 1.036854 & -3.41 & -4.14 & -4.60 & -4.89 & -6.06 & -6.17 & -5.42 & -5.73 & -5.83 \\
XX Cen       & 1.039548 & -3.22 & -3.93 & -4.36 & -4.68 & -5.84 & -5.94 & -5.20 & -5.50 & -5.60 \\
V340 Nor     & 1.052579 & -3.10 & -3.94 & -4.42 & -4.73 & -5.97 & -6.14 & -5.36 & -5.70 & -5.82 \\
UU Mus       & 1.065819 & -3.14 & -3.89 & -4.36 & -4.68 & -5.90 & -6.01 & -5.29 & -5.60 & -5.72 \\
U Nor        & 1.101875 & -3.26 & -4.01 & -4.52 & -4.80 & -6.02 & -6.13 & -5.40 & -5.70 & -5.80 \\
SU Cru       & 1.1088   & -3.47 & -4.30 & -4.88 & -5.23 & -6.67 & -6.75 & -5.99 & -6.53 & -6.66 \\
BN Pup       & 1.135867 & -3.64 & -4.42 & -4.89 & -5.23 & -6.48 & -6.60 & -5.79 & -6.11 & -6.22 \\
TT Aql       & 1.138459 & -3.43 & -4.30 & -4.80 & -5.15 & -6.48 & -6.64 & -5.72 & -6.05 & -6.15 \\
LS Pup       & 1.150646 & -3.63 & -4.40 & -4.87 & -5.20 & -6.44 & -6.55 & -5.76 & -6.09 & -6.19 \\
VW Cen       & 1.177138 & -2.94 & -3.87 & -4.43 & -4.83 & -6.31 & -6.46 & -5.55 & -5.95 & -6.10 \\
X Cyg        & 1.214482 & -3.76 & -4.66 & -5.17 & -5.53 & -6.88 & -7.07 & -6.14 & -6.48 & -6.60 \\
CD Cyg       & 1.232334 & -3.87 & -4.68 & -5.19 & -5.50 & -6.78 & -6.91 & -6.13 & -6.45 & -6.55 \\
Y Oph        & 1.233609 & -3.89 & -4.62 & -5.16 & -5.43 & -6.70 & -6.77 & -5.96 & -6.21 & -6.28 \\
SZ Aql       & 1.234029 & -3.90 & -4.79 & -5.33 & -5.68 & -7.05 & -7.22 & -6.30 & -6.63 & -6.74 \\
VY Car       & 1.276818 & -3.69 & -4.61 & -5.13 & -5.51 & -6.89 & -7.09 & -6.13 & -6.49 & -6.61 \\
RU Sct       & 1.29448  & -3.94 & -4.69 & -5.26 & -5.51 & -6.79 & -6.87 & -6.09 & -6.38 & -6.46 \\
RY Sco       & 1.307927 & -3.91 & -4.65 & -5.19 & -5.48 & -6.76 & -6.84 & -6.09 & -6.38 & -6.49 \\
RZ Vel       & 1.309564 & -3.83 & -4.66 & -5.14 & -5.51 & -6.82 & -6.96 & -6.13 & -6.47 & -6.59 \\
WZ Sgr       & 1.339443 & -3.63 & -4.60 & -5.16 & -5.55 & -7.03 & -7.21 & -6.30 & -6.70 & -6.84 \\
WZ Car       & 1.361977 & -3.83 & -4.61 & -5.09 & -5.43 & -6.70 & -6.81 & -6.08 & -6.42 & -6.54 \\
SW Vel       & 1.370016 & -4.07 & -4.88 & -5.37 & -5.73 & -7.05 & -7.17 & -6.34 & -6.68 & -6.80 \\
T Mon        & 1.431915 & -4.18 & -5.17 & -5.68 & -6.08 & -7.51 & -7.73 & -6.75 & -7.14 & -7.27 \\
RY Vel       & 1.449158 & -4.32 & -5.14 & -5.65 & -5.99 & -7.32 & -7.44 & -6.62 & -6.92 & -7.04 \\
AQ Pup       & 1.478624 & -4.56 & -5.39 & -5.93 & -6.28 & -7.67 & -7.78 & -6.85 & -7.20 & -7.31 \\
KN Cen       & 1.531857 & -4.79 & -5.59 & -6.14 & -6.43 & -7.74 & -7.86 & -7.15 & -7.53 & -7.67 \\
$\ell$ Car   & 1.550816 & -4.11 & -5.22 &       & -6.21 & -7.74 & -8.03 & -6.91 & -7.33 & -7.46 \\
U Car        & 1.588970 & -4.51 & -5.43 & -5.94 & -6.32 & -7.71 & -7.89 & -6.97 & -7.32 & -7.45 \\
RS Pup       & 1.617420 & -4.78 & -5.76 & -6.32 & -6.72 & -8.22 & -8.40 & -7.37 & -7.74 & -7.87 \\
SV Vul       & 1.652569 & -4.97 & -5.97 & -6.53 & -6.90 & -8.35 & -8.58 & -7.52 & -7.86 & -7.96 \\
             &          &       &       &       &       &       &       &       &       &       \\
      \hline
    \end{tabular}
    \label{tab:absmag}
 \end{center}
\end{table*}


The adopted Galactic \emph{PL} relations are given in \Tab{tab:plrel}: slope and intercept correspond to the relation 
$M = a \log P +b$. Their domain of validity extends from 0.57 to 1.65 in $\log P$ (3.7 to 45 days). We also define 
reddening-free Wesenheit magnitudes for the V and $\Ic$ bands, and similarly for the B and $\Ic$ bands, as:

\begin{eqnarray}
\Wvi & = & V - 2.55 \ \VI \\
\Wbi & = & B - 1.866 \ \BI
\end{eqnarray}
where the colour coefficients are computed from the adopted reddening law (see \Tab{tab:redlaw}). For the V and $\Ic$
bands, it is the same as the OGLE adopted coefficient \citep{Udalski1999a}. The \emph{PL} relations for these two 
Wesenheit magnitudes, which are less dispersed than the \emph{PL} relations based on standard visible photometric bands,
are also given in \Tab{tab:plrel}, and data are listed in \Tab{tab:absmag}.

\begin{table*}
 \begin{center}
    \caption{Adopted Galactic and LMC \emph{PL} relations: $M = a \log P + b$. 
             Note that intercept error is for the barycenter of data points (no slope error included).}
    \begin{tabular}{lllcclr}
      \hline\hline
             &           &        &                    &                    &          &     \\
      Galaxy & Source    & Band   & slope a            & intercept b        & $\sigma$ & N   \\
      \hline
             &           &        &                    &                    &          &     \\
      MW     & this work & B      & $-2.289 \pm 0.091$ & $-0.936 \pm 0.027$ & 0.207    & 58  \\
             &           & V      & $-2.678 \pm 0.076$ & $-1.275 \pm 0.023$ & 0.173    & 58  \\
             &           & $\Rc$  & $-2.874 \pm 0.084$ & $-1.531 \pm 0.025$ & 0.180    & 54  \\
             &           & $\Ic$  & $-2.980 \pm 0.074$ & $-1.726 \pm 0.022$ & 0.168    & 59  \\
             &           & J      & $-3.194 \pm 0.068$ & $-2.064 \pm 0.020$ & 0.155    & 59  \\
             &           & H      & $-3.328 \pm 0.064$ & $-2.215 \pm 0.019$ & 0.146    & 56  \\
             &           & $\Ks$  & $-3.365 \pm 0.063$ & $-2.282 \pm 0.019$ & 0.144    & 58  \\
             &           & $\Wvi$ & $-3.477 \pm 0.074$ & $-2.414 \pm 0.022$ & 0.168    & 58  \\
             &           & $\Wbi$ & $-3.600 \pm 0.079$ & $-2.401 \pm 0.023$ & 0.178    & 58  \\
      LMC    & OGLE      & B      & $-2.439 \pm 0.046$ & $17.368 \pm 0.009$ & 0.239    & 662 \\
             & this work & B      & $-2.393 \pm 0.040$ & $17.356 \pm 0.010$ & 0.272    & 714 \\
             & OGLE      & V      & $-2.779 \pm 0.031$ & $17.066 \pm 0.006$ & 0.160    & 650 \\
             & this work & V      & $-2.734 \pm 0.029$ & $17.052 \pm 0.007$ & 0.199    & 716 \\
             & this work & $\Rc$  & $-2.742 \pm 0.060$ & $16.697 \pm 0.020$ & 0.185    & 83  \\
             & OGLE      & $\Ic$  & $-2.979 \pm 0.021$ & $16.594 \pm 0.004$ & 0.107    & 662 \\
             & this work & $\Ic$  & $-2.957 \pm 0.020$ & $16.589 \pm 0.005$ & 0.132    & 692 \\
             & Persson   & J      & $-3.153 \pm 0.051$ & $16.336 \pm 0.015$ & 0.140    & 92  \\
             & this work & J      & $-3.139 \pm 0.026$ & $16.273 \pm 0.006$ & 0.128    & 529 \\
             & Persson   & H      & $-3.234 \pm 0.042$ & $16.079 \pm 0.012$ & 0.116    & 92  \\
             & this work & H      & $-3.237 \pm 0.024$ & $16.052 \pm 0.005$ & 0.117    & 529 \\
             & Persson   & $\Ks$  & $-3.281 \pm 0.040$ & $16.051 \pm 0.011$ & 0.108    & 92  \\
             & this work & $\Ks$  & $-3.228 \pm 0.028$ & $15.989 \pm 0.006$ & 0.136    & 529 \\
             & OGLE      & $\Wvi$ & $-3.309 \pm 0.011$ & $15.875 \pm 0.002$ & 0.056    & 671 \\
             & this work & $\Wvi$ & $-3.320 \pm 0.011$ & $15.880 \pm 0.003$ & 0.070    & 686 \\
             & this work & $\Wbi$ & $-3.454 \pm 0.011$ & $15.928 \pm 0.003$ & 0.076    & 688 \\
      \hline
    \end{tabular}
    \label{tab:plrel}
 \end{center}
\end{table*}

\begin{figure*}
  \begin{center}
    \includegraphics[width=18cm]{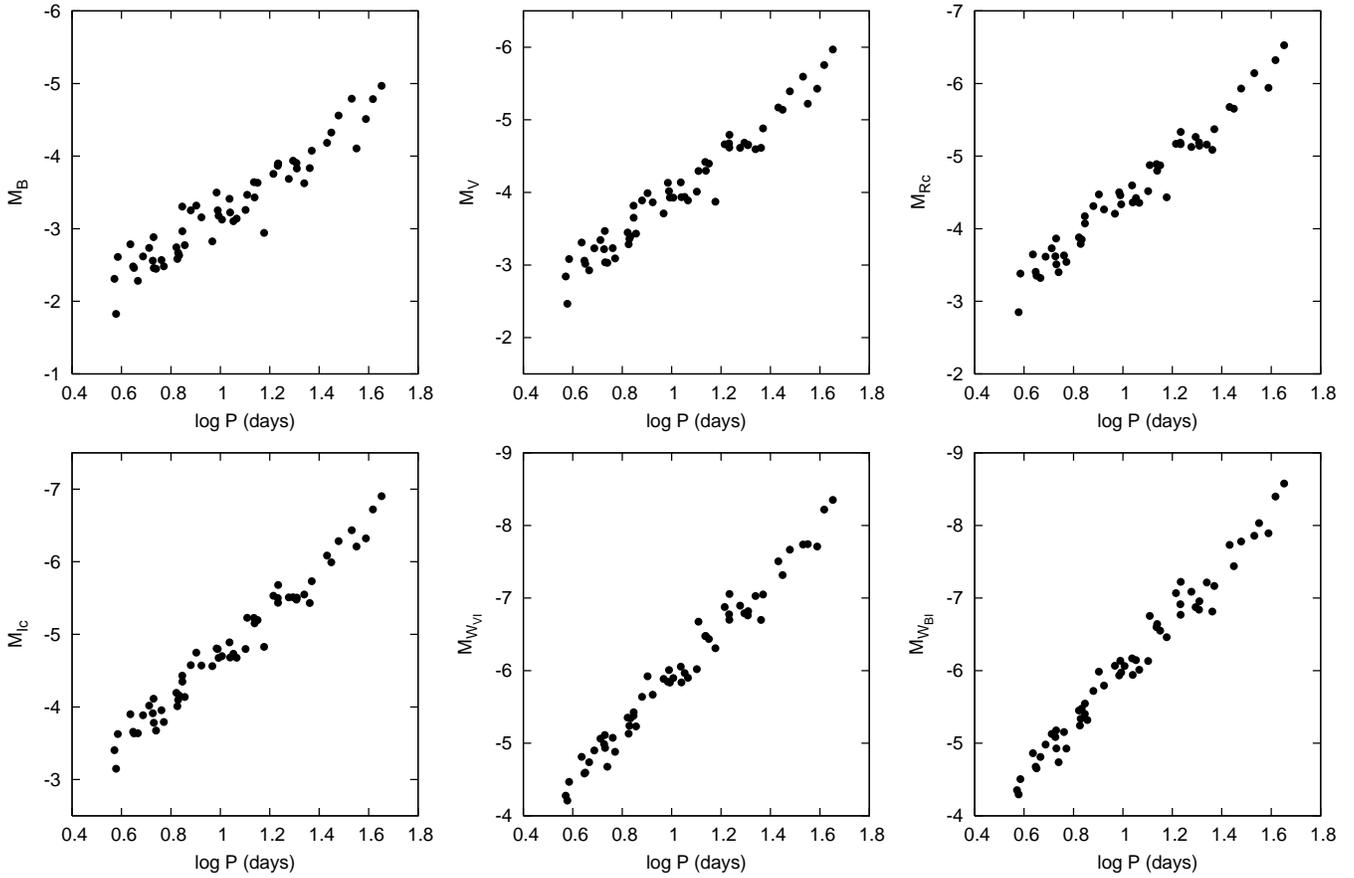}
    \caption{Adopted Galactic \emph{PL} relations in optical bands.}
    \label{fig:optpl}
  \end{center}
\end{figure*}

\begin{figure*}
  \begin{center}
    \includegraphics[width=18cm]{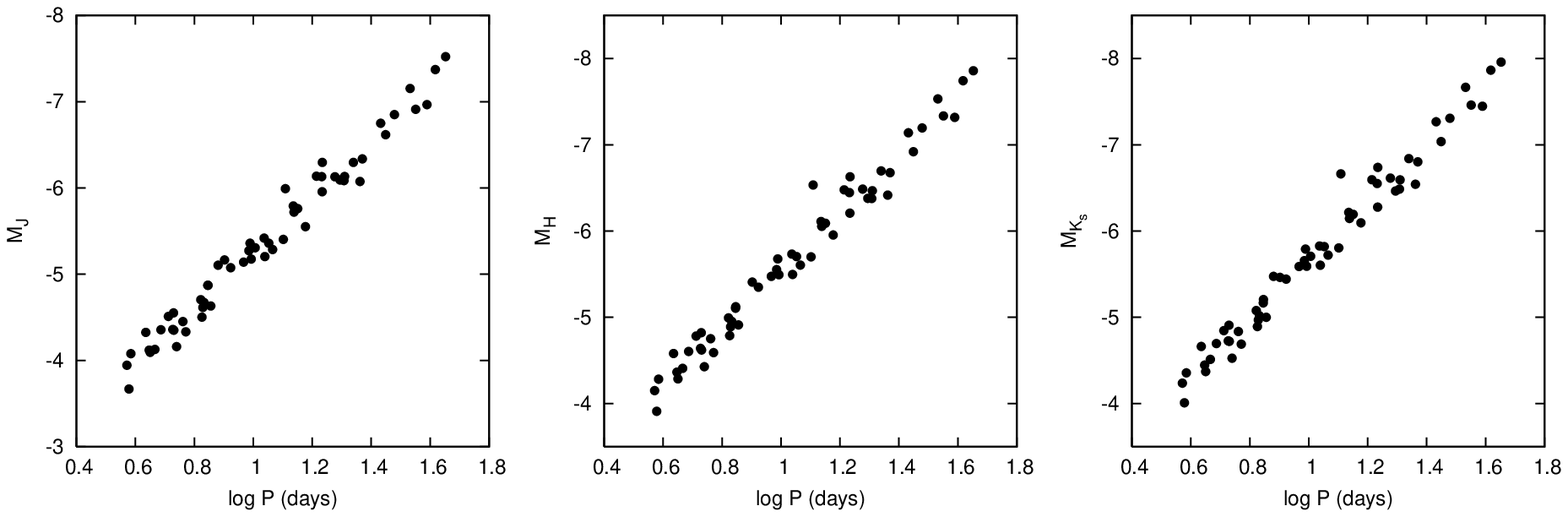}
    \caption{Adopted Galactic \emph{PL} relations in near-infrared bands.}
    \label{fig:irpl}
  \end{center}
\end{figure*}

\Fig{fig:optpl} and \Fig{fig:irpl} display the adopted \emph{PL} relations in the optical bands (B V $\Rc$ $\Ic$ $\Wvi$ and 
$\Wbi$) and in the near-infrared (J H $\Ks$), respectively.

\section{Comparison of the Galactic and LMC \emph{PL} relations} \label{sec:comparison}

We will now compare our new Galactic \emph{PL} relations to the LMC \emph{PL} relations. The choice of the LMC \emph{PL} 
relations in the visible is quite obvious, given the quality of the OGLE sample. The precise values of the slopes 
and intercepts of these relations depend on the adopted reddening corrections and on rejection of some outliers and 
of Cepheids with all kinds of imaginable problems (see \citet{Kanbur2003}), as described in \citet{Fouque2003}. 
For simplicity, we will compare our results to the original OGLE relations from \citet{Udalski1999a}, as updated 
on the OGLE web site\footnote{ftp://sirius.astrouw.edu.pl/ogle/ogle2/var\_stars/lmc/cep/catalog/ \\ README.PL}: see
\Tab{tab:plrel}.

In the infrared, we will use both the \citet{Persson2004} relations based on 92 Cepheids with well-sampled light
curves (22 phase points per star on average), and an OGLE sample of 529 Cepheids ($0.4 < \log P < 1.5$) with 2MASS 
photometry from \citet{Soszynski2005}, where the visible light curves have been used to transform the single-phase 
2MASS measurements into mean magnitudes. We assume that LCO and 2MASS photometry are approximately in the same 
photometric system, which is confirmed by a direct comparison of 18 common stars, which shows an unsignificant shift 
of $-0.02 \pm 0.01$ (in the sense Soszy\'nski minus Persson) in all three bands. \emph{PL} relations are given in \Tab{tab:plrel}.

In order to test for a possible change of slope at about a period of 10 days, as advocated for instance by 
\citet{Tammann2003} and \citet{Sandage2004}, we enlarge the OGLE sample with 173 Cepheids from various sources, 
among which 68 have periods larger than 10 days. We are aware that adding these measurements from different 
photometric systems and with different assumed extinctions may well lead to the kind of systematic effects we want 
to test, so that, in our opinion, a superior way of definitively settling this important question will have to await
for the publication of the OGLE shallow survey of LMC Cepheids. Optical photometry mainly comes from 
\citet{Sebo2002}. We assume the same reddening law as for the Galactic Cepheids. Extinction values for this 
additional sample are adopted from \citet{Sandage2004} Table~1 (85 cases) and \citet{Gieren1998} Table~8 (KMS SW-341) 
when available, or assumed to be 0.1 if not (77 cases). The resulting \emph{PL} relations in B V $\Rc$ $\Ic$ are listed 
in \Tab{tab:plrel}. Six Cepheids with $\log P > 1.8$ have been a priori excluded from the sample. The final sample sizes 
after rejection of outliers are also listed in that Table. Clearly, the new relations have a larger dispersion than the OGLE ones, 
probably because of a lower quality photometry and less accurate extinction values in the additional data. However, the slopes 
and intercepts are compatible within $1 \  \sigma$ with the OGLE ones. The \emph{PL} relations for the two Wesenheit magnitudes are 
also given in that Table. A comparison of these \emph{PL} relations with original OGLE ones shows that adding long-period Cepheids 
does not change the \emph{PL} relations significantly.

\begin{figure*}
  \begin{center}
    \includegraphics[width=14cm]{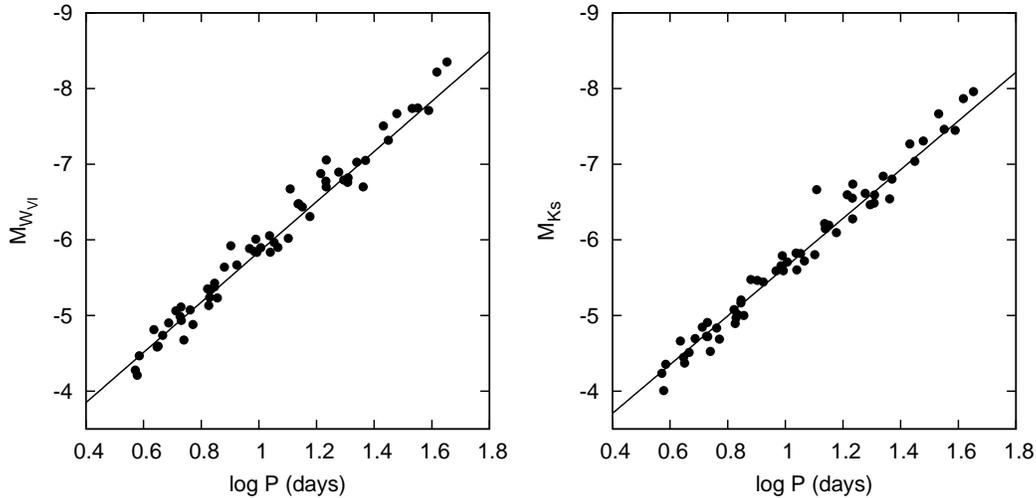}
    \caption{Galactic \emph{PL} relations in $\Wvi$ and $\Ks$ bands, superimposed with LMC Ogle relations shifted by 
    a magnitude offset of 18.40.}
    \label{fig:lmcpl}
  \end{center}
\end{figure*}

Comparison with the Galactic \emph{PL} relations shows a general good agreement, except perhaps in $\Wvi$ and $\Wbi$, where the 
Galactic slopes are slightly steeper. Small remaining differences in all bands may be due to the adopted slope of the \emph{p\,(P)} 
relation, 
which is still uncertain. Good agreement can be judged in $\Wvi$ and $\Ks$ from \Fig{fig:lmcpl}, which displays Galactic data points 
together with LMC \emph{PL} relations (from this work), using a magnitude offset of 18.40. This offset is the sum of the LMC distance 
modulus and any possible metallicity correction. As the same offset seems to work both in $\Wvi$ and $\Ks$, possible metallicity 
corrections must be similar in both bands: for instance, negligible metallicity correction in K and a correction of 0.13 mag in $\Wvi$,
as advocated by \citet{vanLeeuwen2007}, is in marginal disagreement with our findings. Moreover, as most published metallicity 
corrections are small and negative (which if true will make the more metal-poor LMC Cepheids intrinsically fainter, at a given period, 
than their Milky Way counterparts, implying a nearer LMC), and taking into account the dispersion of the Galactic data points, 18.50 
appears to be an upper limit to the LMC distance modulus from our present study.

\section{Summary and conclusion} \label{sec:discussion}

The recent publication of accurate HST parallaxes of Galactic Cepheids has prompted new studies of the universality of the
\emph{PL} relations. However, the small number of measured Cepheids with this technique (10) does not allow an accurate
determination of the slope of the Galactic \emph{PL} relations. Similarly, the recent publication of revised Hipparcos
parallaxes of Cepheids has the same limitation, because few Cepheids have a parallax accuracy better than 30\%, and most
of them are first overtone pulsators.

The main effort of the present study has been to increase the size of the sample of Galactic Cepheids with known parallaxes,
using mainly the infrared surface brightness technique. Seventy Galactic Cepheids have now been measured by this method, among
which 54 are suitable to calibrate the Galactic \emph{PL} relations. We have chosen a recent determination of the \emph{p\,(P)} 
relation based on hydrodynamical models to derive IRSB parallaxes. We verified that the HST parallaxes were compatible in the mean 
with these IRSB parallaxes. Adding five stars from other techniques, among which the interferometric Baade-Wesselink one, we are 
able to show that the Galactic slopes do not significantly differ from the corresponding slopes in the LMC for a given photometric 
band, from B to K.

This important result shows that applying the well determined LMC slopes to galaxies of different metallicity contents is
warranted. Possible metallicity effects in the zero-point of the relations are not studied in the present work, and may still
prevent a precise determination of galaxy distances using Cepheids. In the case of the LMC, the true distance modulus (corrected for
metallicity effects) appears to be smaller than 18.50.

\begin{acknowledgements}

We express our gratitude to Leonid Berdnikov, John Caldwell, Michael Feast, Shashi Kanbur, 
Dave Laney, Chow Choong Ngeow, and Igor Soszy\'nski for communicating unpublished data.

This publication makes use of data products from the 2MASS project, as well as the SIMBAD 
database, Aladin and Vizier catalogue operation tools (CDS Strasbourg, France). 
The Two Micron All Sky Survey is a joint project of the University of Massachusetts and the
Infrared Processing and Analysis Center/California Institute of Technology, funded by 
the National Aeronautics and Space Administration and the National Science Foundation.

PA acknowledges the ALFA/LENAC network under the European Commission ALFA grant programme
for financing her training at the Laboratoire d'Astrophysique de Toulouse.

This material is based in part upon work by TGB while serving at the National Science Foundation. 
Any opinions, findings, and conclusions or recommendations expressed in this material are those 
of the authors and do not necessarily reflect the views of the National Science Foundation.

WG acknowledges support for this work from the Chilean Center for Astrophysics FONDAP 15010003.

PF dedicates this work to Alain Milsztajn, who contributed to a better knowledge of Cepheids
through the EROS survey, and passed away on June 28, 2007.

\end{acknowledgements}

\bibliographystyle{aa}
\bibliography{8187}
\end{document}